\documentclass[graybox]{svmult}

\usepackage{type1cm}        
\usepackage{makeidx}         
\usepackage{graphicx}        
\usepackage{multicol}        
\usepackage[bottom]{footmisc}

\usepackage{newtxtext}       %
\usepackage[varvw]{newtxmath}       
\usepackage[T1]{fontenc}

\makeindex             

\begin{document}

\title*{Towards Intelligent Active Particles}
\author{Hartmut L\"owen, Benno Liebchen}
\institute{Hartmut L\"owen \at Institut für Theoretische Physik II: Weiche Materie, Heinrich-Heine Universit\"at, Universit\"atsstraße 1,  40225 D\"usseldorf, 
\\
\email{hlowen@hhu.de}
\and Benno Liebchen \at Technische Universit\"at Darmstadt, Hochschulstr. 8, 64289 Darmstadt, \\
\email{benno.liebchen@pkm.tu-darmstadt.de}}
%
%
\maketitle

\abstract{-- In this book chapter we describe recent applications of artificial intelligence and in particular machine learning to active matter systems. Active matter is composed of agents, or particles, that are capable of propelling themselves. While biological agents like bacteria, fish or birds naturally possess a certain degree of ``intelligence'', synthetic active particles like 
colloidal microswimmers and electronic robots can be equipped with different levels of artificial intelligence, 
either internally (as for robots) or via a dynamic external control system. This book chapter briefly discusses existing approaches to make synthetic particles increasingly ``intelligent'' and 
then focuses on the usage of machine learning to approach navigation and communication problems of active particles. Basic questions are how to steer a single active agent through a complex environment to reach or discover a target in an optimal way and how active particles need to 
cooperate to efficiently collect a distribution of targets (e.g. nutrients or toxins) from their 
complex environment.}

\section{Introduction}
\label{sec:Introduction}
The past few years have seen a tremendous development in combing active matter \cite{mestre2022colloidal} with artificial intelligence (AI) \cite{cichos2020machine}. Active matter typically comprises agents (or `particles'') like bacteria, algae, vibrated granulates or synthetic colloidal microswimmers that convert energy from their environment into directed mechanical motion \cite{bechinger2016active, gompper20202020}. These agents are out of equilibrium, in qualitative difference to passive particles. Since more than two decades \cite{paxton2004catalytic} it is possible to create synthetic active particles in the colloidal regime, i.e. at the micron- or the nano-scale, where they are 
typically embedded in a surrounding liquid \cite{bechinger2016active}, or sometimes in a complex plasma \cite{nosenko2020active}. Activity then implies that the particles are actively moving (self-propelling) relative to the surrounding medium and are, unlike passive particles, not just advected by it. Accordingly, colloidal active particles are also viewed as 
micro- and nano-motors that do not require any movable parts and that can transport cargo at much smaller scales than conventional motors. 
They are also frequently discussed for their potential to perform useful tasks in the future such as delivering drugs to cancer cells, performing microsurgery, and collecting microplastics. However, this would either require to externally steer each individual particle in a suitable way, which is not easy if they are for example deep inside the human body,
or to equip the particles with some degree of artificial intelligence that allows them to navigate
autonomously. 
To date it remains as a persistent experimental challenge to 
directly integrate the required hardware for artifical intelligence into active colloidal particles \cite{nasiri2023optimal,cichos2023artificial} which are far too small to equip them with 
sensors, actuators and information processing units. Thus, other, simpler, approaches are required, some of which we discuss in the following. 
\\Figure 1 shows particles of increasing complexity that are 
equipped with seven different levels of sophistication or ``intelligence''. Here the individual levels can be viewed as steps between an ordinary particle without any intelligence and a complete sensor-processor-actuator system that can perceive information and can perform computations to determine actions that are executed by the actuator. 
Concretely, panel a) represents a passive particle with no ``intelligence'' at all. This particle is randomly kicked by a Newtonian solvent (such as water) leading to Brownian motion. Such particles are the 
subject of traditional colloid science which has started more than 100 years ago with famous contributions from Einstein, Perrin, Derjaguin, Landau, Verwey and Overbeek and has been brought to maturity by using almost monodisperse suspensions even at high particle density as ideal classical model systems for equilibrium phase transitions \cite{yethiraj2007tunable}. In b) the particle is susceptible to an external field and can be steered by suitable variations of the field in space and time. Possible fields are an external solvent flow field, a laser-optical field (leading to optical laser-tweezers), an electric or a magnetic field \cite{lowen2013introduction}. Here,  
the particle itself also has no intelligence at all, but the setup allows a human controller to use her/his intelligence to steer the particle in a desired way. 
This can be done based on quick feedback effects (e.g. using optical tweezers) such that the particle motion can be changed at each point in time almost at wish. 
The next level, c), is an active colloidal particle which has internal degrees of freedom and which can self-propel, i.e. it is self-actuated. Such active colloids are also called microswimmers and they are the elementary building blocks of active soft matter. The trajectory of a microswimmer typically has an initial ballistic part (apart from Brownian motion at very short times) which crosses over to long-time Brownian motion albeit with a so-called active diffusion coefficient which typically is orders of magnitude larger than the diffusion coefficient of passive colloids \cite{bechinger2016active}. The intelligence of such microswimmers is rather primitive; it is embodied in the internal motor of the particles which reacts to environmental changes. For instance, it is now well known that active colloids which catalyze a certain chemical reaction on part of their surface only and use a self-created concentration gradient to achieve self-propulsion also react to 
external concentration gradients. Effectively, this reaction is 
similar to the reaction of chemotactic bacteria to nutrient or toxin gradients (albeit based on a very different mechanism) \cite{liebchen2018synthetic}. In this sense, active colloids are self-actuated agents that can sense and react to their environment. 
The situation can be made more complex (see d)) when dispersing the particle into a non-Newtonian solvent (such as a polymer solution). Then the solvent background exhibits memory effects which leads to a non-Markovian dynamics. In some sense, compared to case c), the ``intelligence'' of the system has increased as it now remembers the past (via its environment) and its dynamics is influenced by its history (or memory). In panel e) the particle is equipped with a sensor that is directly coupled to the self-propulsion mechanism such that the particle reacts to its environment. That is the particle measures an external stimulus and transforms it instantaneously into motion. In contrast to case c) where the particle also reacts to its environment, in e) the sensation can be more general and can for example involve a certain vision cone \cite{lavergne2019group}.
If this is augmented with an internal memory storing past sensations (panel f), the particle is more ``intelligent'' in the sense that it can not only react to its environment in a way that depends on its present sensory input but that can also depend on the past, i.e. on the particle's experience. 
Finally, panel g) shows an active particle that is equipped with a sensor and a processor that serves as an internal brain. Such a particle represents a full actuator-sensor-processor system.
Such systems can be realized in the form of programmable electronic robots at centimeter if not millimeter sizes, but not yet at the microscale. However, while we are not yet able to synthetically create intelligent micro-devices with the properties indicated in panel g), 
nature has already realized them in the form of bacteria and other microorganisms. 

\clearpage
\begin{figure} [h]
    \centering
    \includegraphics[width=\textwidth,height=0.5\textheight,keepaspectratio]{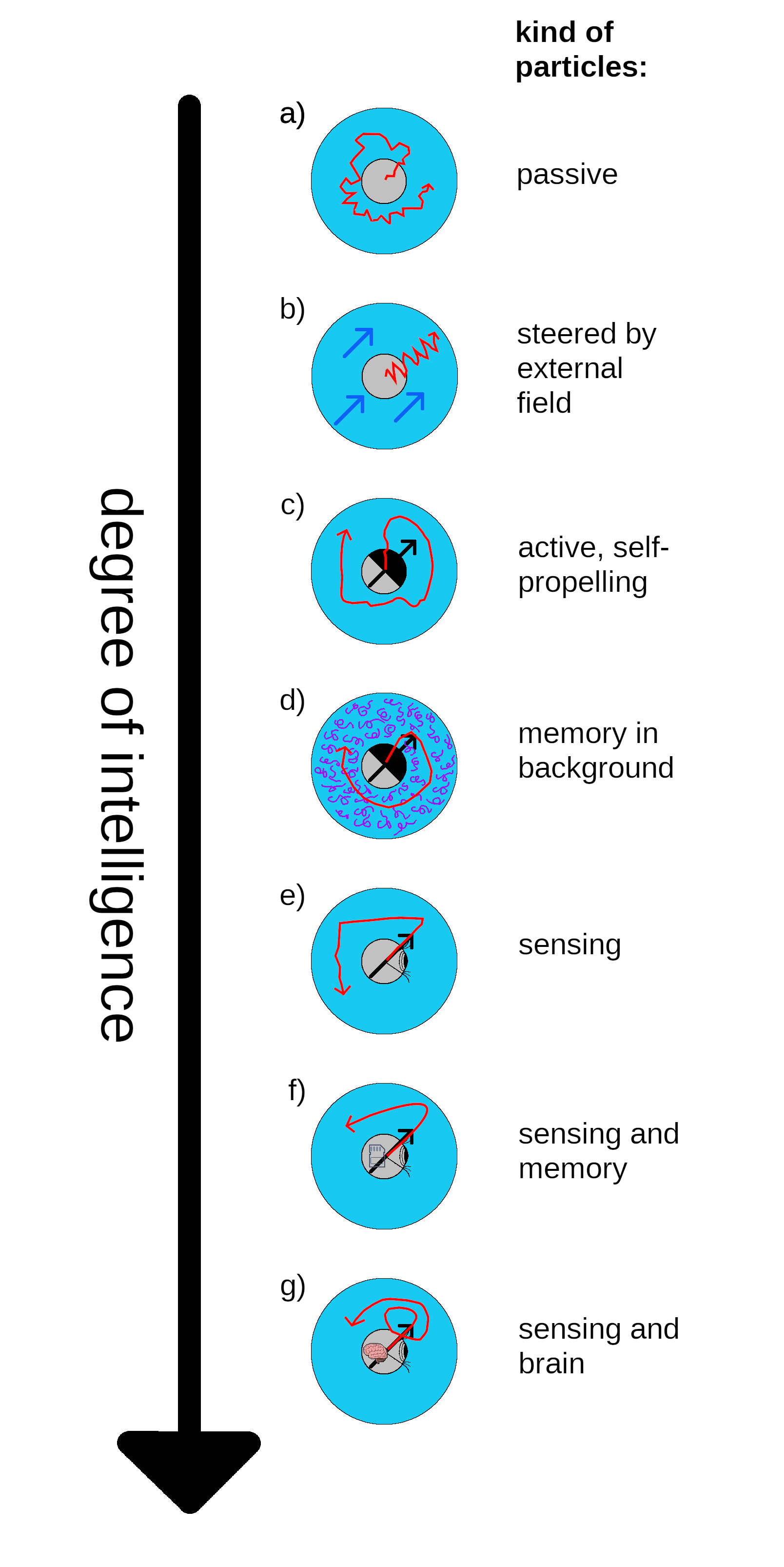}
    \label{fig:figure1}
    \caption{Schematic illustration of micron-sized colloidal particles (grey) in a fluid solvent (blue background) with different levels of sophistication or ``intelligence'' (increasing from top to bottom). Typical trajectories are shown in red color: a) Passive particle with no ``intelligence'' at all; the particle is randomly kicked by the solvent leading to Brownian motion. b) Particle susceptible to an external field (flow, optical electric, magnetic etc) indicated by the black arrows such that it can be steered by an ``intelligent'' external controller at wish. c) Active particles with an inner motor (actuator) leading to self-propulsion (black arrow). These colloidal ``microswimmers'' can move in a way that depends on their local environment. This essentially equips them with the ability to sense and react to their environment in a primitive way (e.g. via synthetic taxis).  
    d) Particles in a non-Newtonian (viscoelastic) medium (indicated by blue polymer coils)  exhibit additional memory effects which are induced externally by the background. These particles react also to the history of their environment. 
    e) Particles equipped with an internal sensor, that allows them e.g. to see and react to other particles within a certain vision cone or to ``smell'' and react in a sophisticated way to chemical cues. f) Particle equipped with a sensor (as indicated by a schematic eye) and an internal memory storage (as indicated by a schematic memory chip) allowing it to react to its environment in a way that depends on present and past sensation (experience). g) Particle that is equipped with a full sensor-processor-actor system, where the processor serves as a ``brain'' that can compute and autonomously make decisions on how to move.}
\end{figure}

\noindent  While we are far away from being able to synthetically realize a fully intelligent micro- or nanoparticle (level g)), AI has been applied in the situations b)-e) in various ways. In particular when using external fields lead to actuate motion of particles and a feedback control system \cite{khadka2018active, bauerle2018quorum,lavergne2019group,sprenger2020active} ``intelligent'' behavior can be implemented. Here, the learning process takes place externally in a computer (or some other machine or human). 
Accordingly, one focus of current studies is to steer particles (actuated or/and self-propelled) with external fields that are dynamically adapted to the motion and environment of the particle. 
Alternatively, one can also use small robots as active particles \cite{mijalkov2016engineering}. Then the learning process takes place internally. 
In both cases AI can be applied either to solve learning problems for an \textit{individual} swimmer or for an ensemble of active agents. 
Problems for an individual swimmer can concern e.g. the problem to 
find the optimal route to a food source (e.g. the fastest path) that is located within a complex environment 
or to find the optimal body-shape deformations of a deformable swimmer to optimize the self-propulsion speed.
Here, challenges in determining e.g. optimal navigation or swimming strategies arise 
from the smallness of the swimmers. This is because 
microswimmers are subject to significant fluctuations due to Brownian motion, errors and delays in the steering protocol, changes and fluctuation in the environment; hence they cannot accurately predict the outcome of their navigational maneuvers and this hast to be taken into account for real applications. 
The second problem concerns \textit{interacting active particles} that can in general 
communicate with each other e.g. via self-induced gradients in a phoretic field (e.g. a chemical concentration field, temperature or electric potential) or via light sources and sensors \cite{mijalkov2016engineering}.
Here, artifical intelligence can be used for example to optimize the communication rules among the agents as we exemplarily discuss further below. 
In the future, AI could be also used to optimize communication rules to make the agents cooperate to collectively perform a certain task. 
For macroscopic robots this has been established to a certain extent, but if we have an army of agents at the micrometer level (as required e.g. for a drug delivery task in our human body \cite{patra2013intelligent,alapan2018soft, gu2022artificial}) this is getting enormously difficult due to fluctuations and hydrodynamic interactions. 
In the real living world of biological microswimmers, there are many examples of such concerted collective behavior (that is typically evolutionarily optimized). Examples are sperm cells which sense gradients in the concentration of the chemicals emitted by the egg  
and phagocytes hunting pathogenic bacteria \cite{spehr2003identification,eisenbach2006sperm}. 
In contrast, while synthetic microswimmers can also sense and react to their environment, even via self-produced chemical signals \cite{liebchen2018synthetic,tsang2020roads}, they do not yet cooperate in a way that optimizes their collective behavior to perform useful tasks. 
\\In this book chapter we shall focus on applications of AI on microswimmers, both single and collective. We first focus on reinforcement learning techniques to guide and steer single microswimmers in a complex environment. Then we discuss predator-prey systems and many-body problems which feature applications of AI for communicating agents. Here we emphasize recent applications and do not aim to be exhaustive. Instead, for more exhaustive discussions of applications of artificial intelligence to active matter we would like to guide the reader to the articles 
 \cite{nasiri2023optimal,cichos2023artificial} and to the bookchapter by G. Volpe, and the one by J. Jeggle and R. Wittkowski.

\section{Artificial intelligence applied to a single microswimmer}
\label{sec:2}
An agent that has the ability to mechanically move its limbs can use different strategies to achieve self-propulsion. A necessary condition to achieve a net propulsion at low Reynolds numbers is to break time-reversal symmetry of the mechanical motion pattern. This was exemplified in different model swimmers such as the three-linked-sphere microswimmer of Najafi and Golestanian \cite{najafi2004simple} where three aligned linked spheres move relatively to each other. Another example is the push-me-pull-you swimmer that requires only two spheres that additionally change their size \cite{avron2005pushmepullyou}.
Such model swimmers have been generalized towards circular swimming \cite{ledesma2012circle} and have also been studied in confinement \cite{daddi2018swimming}. 
The three-linked microswimmer can use 
different locomotoric gaits (different shape-deformation patterns), obtained via reinforcement learning, e.g. to optimize self-propulsion speed 
or to steer (navigate) in a desired way 
as recently demonstrated in \cite{tsang2020self,zou2022gait, abdi2023self,qin2023reinforcement}. 
A similar idea can also be applied to mechanical rotations of the swimmer \cite{liu2021mechanical}. 
A second type of problem for which reinforcement learning plays a key role is to learn optimal strategies to find a target of unknown position by chemotaxis such as e.g. a static food source \cite{kaur2023adaptive, goh2022noisy}. Here, an internal learning scheme with a genetic algorithm which leads to sucessfully finding the target was proposed in \cite{hartl2021microswimmers,ramakrishnan2023learning}.
A third typical problem for a single microswimmer that can freely control its self-propulsion direction but not its speed is to find the fastest path from a prescribed starting point A to another prescribed target point B, i.e. the connecting path that leads to the shortest traveling time. Some simple cases can be solved analytically \cite{liebchen2019optimal, daddi2021hydrodynamics} but for more complex environments reinforcement learning techniques       \cite{sutton2018reinforcement} have been applied to find the optimal path, see e.g.  \cite{Yang2018,Yang2019, biferale2019,yang2020micro,xu2021brownian,alageshan2020machine, zhu2022numerical,zhu2022point, nasiri2022reinforcement}. Some works have used tabular Q-learning, for self-thermophoretic active particles \cite{muinos2021reinforcement}, or for learning to navigate optimally inside a Mexican hat potential \cite{schneider2019}. Other works have used actor-critic methods \cite{biferale2019} and also reinforcement learning combined with neural networks \cite{gunnarson2021learning, nasiri2022reinforcement}. In particular, in ref. \cite{nasiri2022reinforcement} a reinforcement learning-based method has been developed to determine {\it asymptotically optimal trajectories} 
which has shown that it is possible to systematically learn the result of an optimal control problem without having to do an explicit calculation.
As an example of such optimal navigation problems we now consider an active particle 
in a prescribed motility landscape, that is in a spatially varying external field that controls the 
self-propulsion speed of the particle. (Such motility patterns can be created e.g. for 
light-driven microswimmers in light-intensity patterns \cite{lozano2016phototaxis}.) 
Suppose that a particle in such a motility pattern can freely choose its self-propulsion direction, but not its speed and let us ask for 
the fastest route that connects a given starting point and a target point 
\cite{monderkamp2022active}. Using reinforcement learning 
and training the agent in different 
motility landscapes allows one to find a navigation strategy that leads to trajectories that closely approximate the exact optimal path, although there is typically no guarantee that the resulting path is really optimal in all cases \cite{monderkamp2022active}. In particular, there is the risk that the reinforcement learning approach converges to some local optimum rather to the global one, which can 
be largely avoided by using a suitable on-policy approach 
\cite{nasiri2022reinforcement}.
Exemplaric results are shown in Figure 2 where the learned trajectory gets closer and closer to the optimal one as training proceeds. 

\begin{figure}
    \centering
    \includegraphics[width=\linewidth]{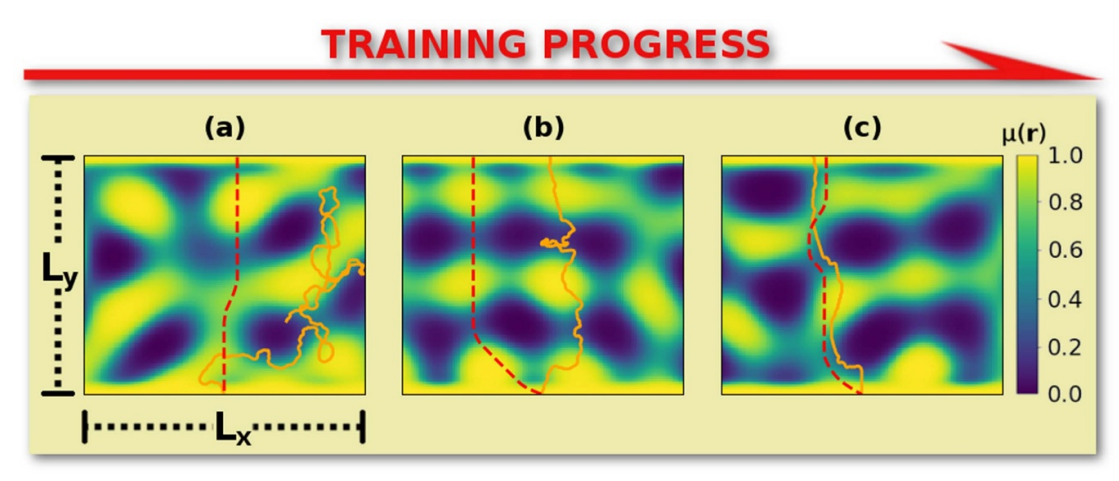}
    \caption{Three typical trajectories (orange solid lines) in three different motility fields $\mu ({\bf r})$ over the course of the training procedure of a Q-learning active Brownian particle. The particle’s objective is to cross the box of lateral sizes $L_x$ and $L_y$ from bottom to top as fast as possible. The motility fields are generated with the help of modified isotropic Gaussian random waves. The probabilities $\epsilon$ to perform random actions are (a) $\epsilon$ = 1.0, (b) $\epsilon \approx$ 0.58, (c)~$\epsilon \approx$~0. The red dashed lines highlight the optimal trajectories obtained with Dijkstras algorithm \cite{dijkstra1959note}. Motility fields are generated as modified isotropic Gaussian random waves. In the horizontal direction reflecting boundary conditions are employed. From ref. \cite{monderkamp2022active}.}
\label{fig:Monderkamp}
\end{figure}

\section{Artificial intelligence applied to predator-prey-like systems}
\label{sec:3}
A somewhat more complex problem occurs if the active particles does not simply navigate towards a stationary target but if the target moves, either randomly or by reacting to the active particle. The latter situation resembles a predator-prey problem where the predator hunts the prey and where the prey tries to escape from the predator. Such problems are very common in nature at the macroscale, but there are also popular examples at the microscale such as, e.g. 
a macrophage chasing a pathogenic bacteria. Recently, such predator-prey-like systems have also been artificially realized in the context of micron-sized particles, see e.g. \cite{meredith2020predator}. (Corresponding models mainly for chemotactic coupling have been explored theoretically \cite{sengupta2011chemotactic, schwarzendahl2021barrier,liebchen2020modeling}.) It is interesting to consider a situation where both the predator and the prey are trained in a reinforcement learning framework and learn by self-play. This has been considered recently for simple lattice models \cite{wang2020reinforcement} relevant for ecology.
Recently, ref.~\cite{gerhard2021hunting} has used reinforcement learning also to explore a predator-prey problem at the colloidal scale. In this study a single predator is surrounded by a gas of prey particles and the prey learns to escape the predator. Different sensing situations are discussed in this work, leading to different strategies. This model can be extended to a group of predators exploring group chasing strategies which have been considered in the context of active particles \cite{janosov2017group}. In ref.~\cite{xu2021brownian} an interesting application has been proposed where a swarm of predators is trained to find a passive but Brownian cargo particle (``prey'') in a maze-like environment. Also the control of particle groups by chasers or controllers (shep-dog in the macroscopic world) \cite{ranganathan2022optimal} are interesting problems where learning strategies will be fruitful.

\section{Artificial intelligence applied to groups of active particles}
We now turn to situations involving a group of particles which are communicating with each other \cite{zampetaki2021collective,ziepke2022multi} and where the group as a whole can learn strategies for different purposes \cite{panait2005cooperative}. 
In particular, collective learning techniques have been applied to active particles for instance to make them learn to swarm and flock \cite{durve2020learning}, to cluster \cite{Speck2014} and to rotate a rigid rod \cite{tovey2023environmental}. Moreover, in groups of fish, optimal swimming patterns in a complex flow field that is collectively generated by the members of the group themselves can also be learned based on deep reinforcement learning \cite{verma2018efficient}. In this case the impact of fluid vortices is key to understand the collective behavior. 
\\Let us now discuss a specific example where a group of communicating agents learns to cooperate in a way that enhances their nutrient consumption \cite{GrauerPREPoptimizing}. 
Concretely, we consider a group of active agents each of which can sense the local food gradient (typical for chemotactic microorganisms) and 
consumes nutrients (or remove toxins) at a certain rate in the vicinity of its momentaneous position. The different agents communicate via chemical signaling (quorum sensing); that is each agents emits certain chemicals which spread out by diffusion and other agents can sense (measure) the gradient in the chemical concentration (Figure 3). That is, overall, each agent can either greedily follow the nutrient concentration gradient which bears the risk of ending up in some small local concentration maximum, or it can try to find a suitable compromise between following the food concentration gradient and the gradient of signaling molecules. The latter allows the agents for example to benefit from other agents that have already found regions of higher nutrient concentration. The key question is then: what is the optimal compromise? 
To find a suitable compromise, one can use reinforcement learning combined with neural networks. This leads to three qualitatively different motion patterns of the agents \cite{GrauerPREPoptimizing}, which are shown in Figure 4. 
Panel a) shows a scenario where the particles cluster in a region of high nutrient concentration. In panel b) particles behave adaptively; they move towards locations of high particle density when they are in regions of low food concentration and they tend to avoid others when they are in regions of high food concentration. Panel 
c) show a case where particles move away from each other and spread over the full system. Figure 5 shows a state diagram where one can see which of these three strategies leads to a higher average nutrient consumption depending on the 
consumption rate of the agents and the agent density (in suitably reduced units). 
This example shows that machine learning can be used to coordinate and direct collective behavior in a way that allows a group of agents to approach a common goal. 

\begin{figure}
    \centering
    \includegraphics[width=\textwidth]{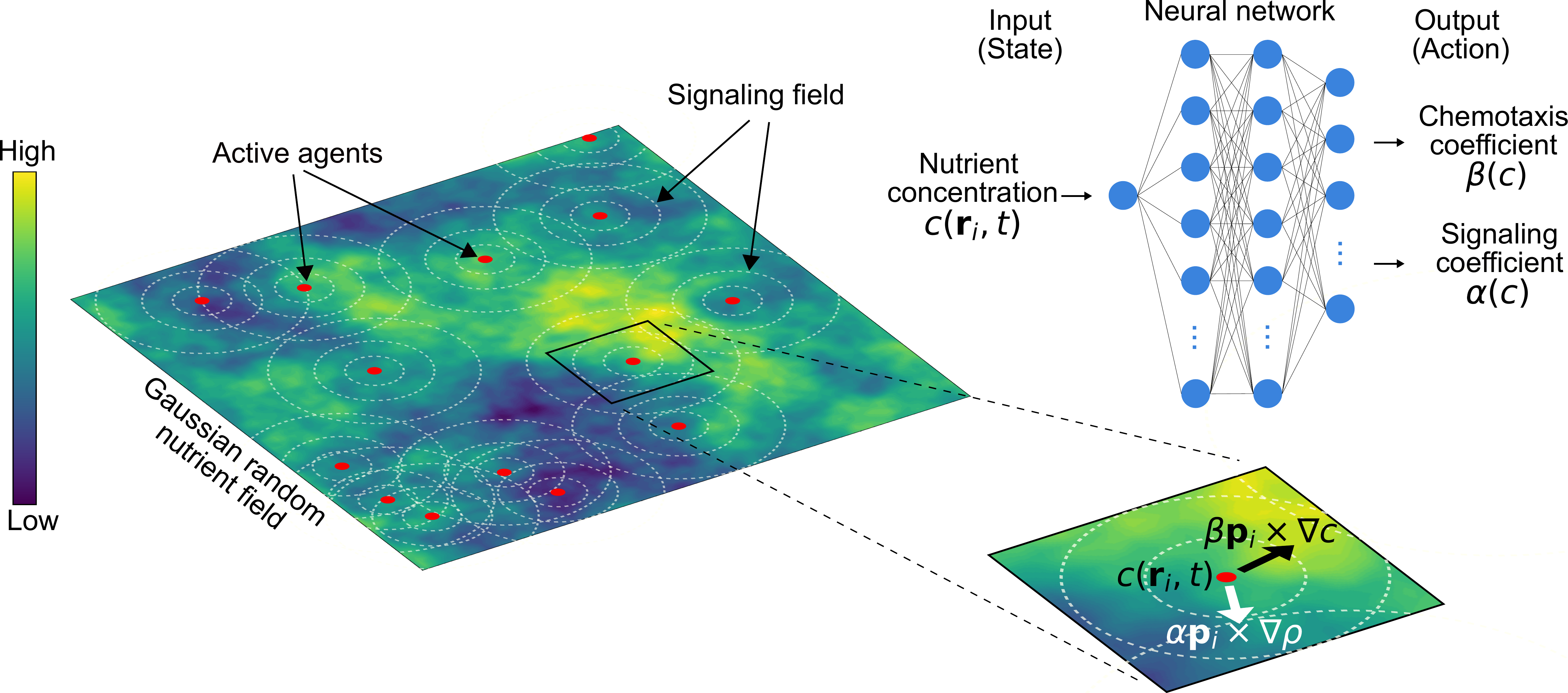}
    \caption{Left: Schematic illustration of active agents (red dots) that 
move in a nutrient concentration field (background color indicates nutrient concentration) and communicate with each other to coordinate their motion. The agents communicate 
by producing quorum sensing molecules (white circles illustrate iso-concentration lines of the signaling molecules due to each agent).
Right: 
Each particle feeds the nutrient concentration at its current position into a neural network. The network helps predicting coupling coefficients 
$\beta$ and $\alpha$ that determine to which extent agents greedily move up nutrient concentration gradients $\nabla c$ and to which extent they follow the concentration gradient of the 
signaling molecules $\nabla \rho$, respectively. From ref. \cite{GrauerPREPoptimizing}.}
    \label{fig:GrauerFig2}
\end{figure}

\begin{figure}
    \centering
    \includegraphics[width=\textwidth]{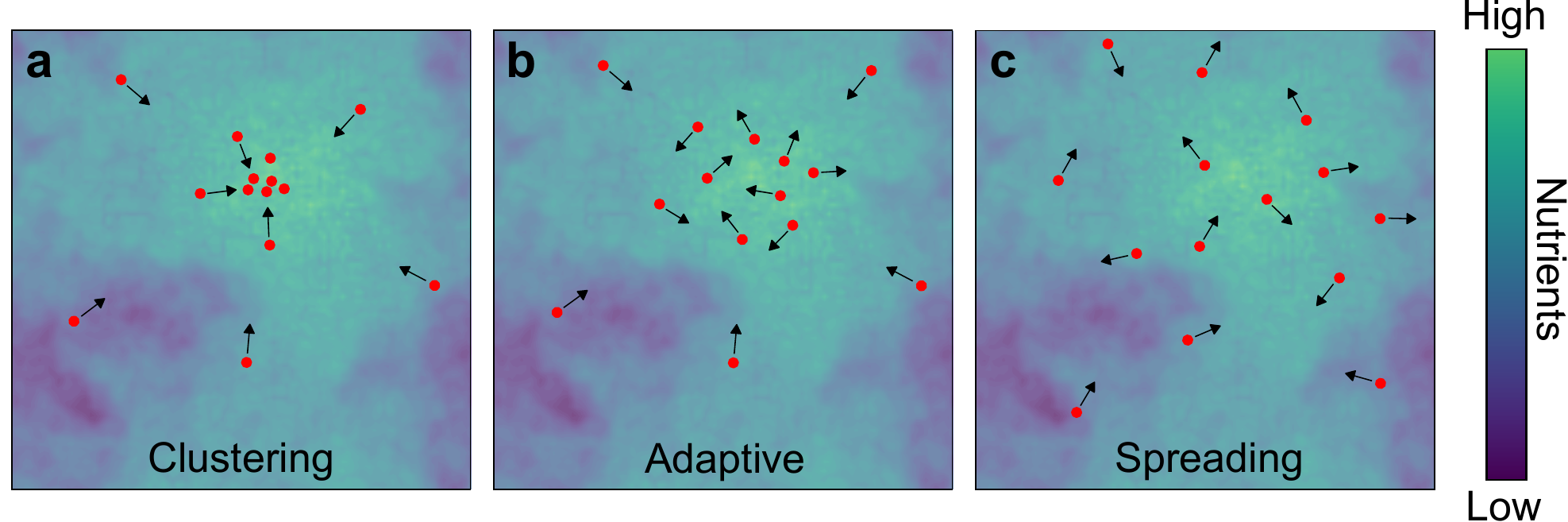}
    \caption{Illustration of the collectively learned motion patterns.
    \textbf{a} clustering strategy: all agents behave cooperatively and swarm together. \textbf{b} adaptive strategy: the agents either follow others or avoid them depending on the local nutrient concentration. \textbf{c} spreading strategy: the agents avoid each other and spread out over the entire domain. From ref. \cite{GrauerPREPoptimizing}.}
    \label{fig:GrauerFig1}
\end{figure}

\begin{figure}
    \centering
    \includegraphics[width=\textwidth]{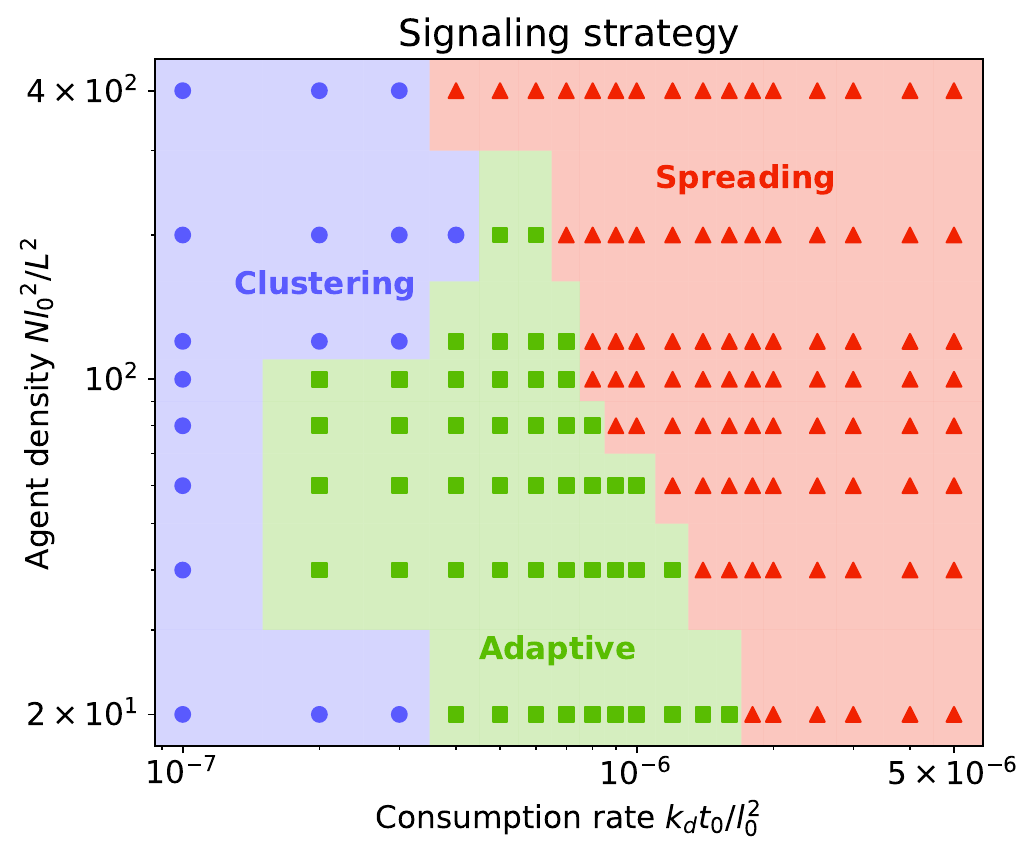}
    \caption{State diagram of learned nutrient collection strategies. The active agents employ different strategies depending on the reduced agent density $N l_0^2/L^2$ and the reduced nutrient consumption rate $k_0 t_0/l_0^2$. Clustering strategy (blue dots): the agents behave cooperatively and aggregate; adaptive strategy (green dots): the agents either approach others or avoid them, depending on the local nutrient concentration; spreading strategy (red dots): the agents avoid each other at all nutrient concentrations. See~\cite{GrauerPREPoptimizing} for details.}
    \label{fig:GrauerFig3}
\end{figure}

\begin{figure}
    \centering
    \includegraphics[width=\textwidth]{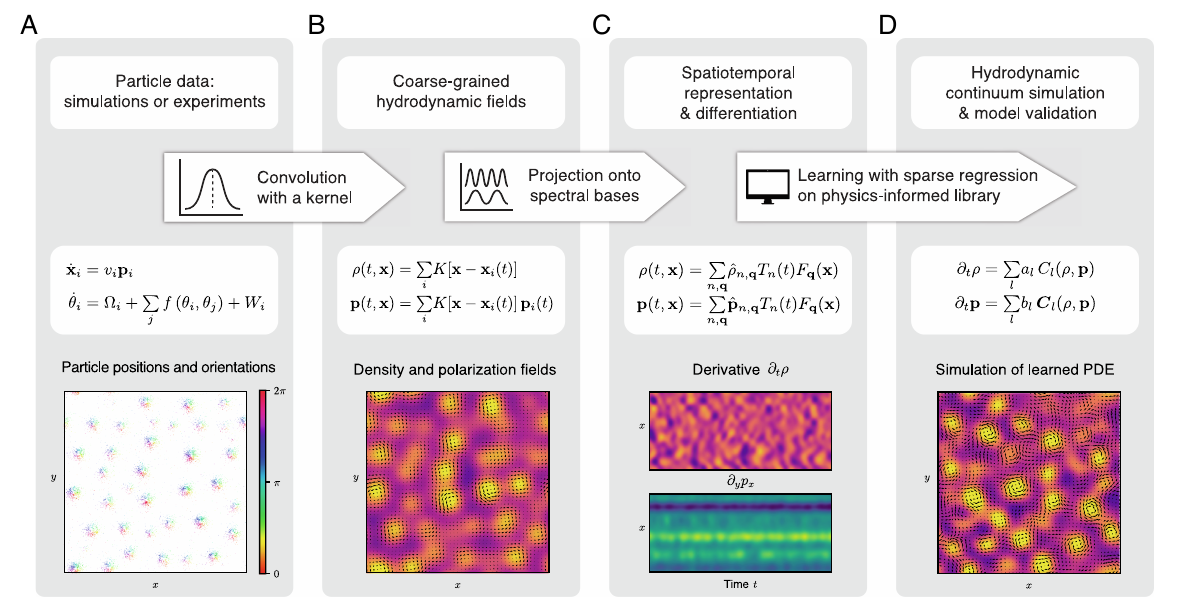}
    \caption{Illustration of the learning approach in ref.~\cite{Supekar_2023}. A.) Inputs are trajectories, e.g. of particle positions ${\bf x}_i(t)$ and orientation angles $\theta_i(t)$, $i=1,2,...,N$, from experiments or particle based simulations. B.) Continuous hydrodynamic fields, such as density $\rho({\bf x},t)$ and polarization ${\bf p}({\bf x},t)$ are obtained via spatial kernal coarse-graining of the discrete microscopic variables. C.) The coarse-grained fields are sampled on a grid and projected onto spectral basis functions. Then, spectral filtering is used to smoothen the fields before calculating derivatives. D.) From these derivatives a library of candidate terms is constructed (consistent with prior knowledge about conservation laws and symmetries) and a sparse regression algorithm is used to determine phenomenological coefficients to obtain a sparse hydrodynamic model. From ref. \cite{Supekar_2023}.}
    \label{fig:supekarDunkelPnas2023}
\end{figure}

\clearpage
\section{Further applications}
There are many other applications of machine learning to active matter physics. For brevity we only sketch a few of them here. For instance, machine learning is currently used to identify phase transitions in active matter, which is a challenging problem even in relatively simple active matter systems due to the strong fluctuations in these systems.
However, recent works have exemplified the possibility to identify phase transitions across a wide range of physical systems, based on various machine learning techniques hinging on supervised learning \cite{carrasquilla2017machine}, unsupervised learning \cite{rodriguez2019identifying,wetzel2017unsupervised}
or combinations thereof \cite{van2017learning}. Very recently, such (or similar) ideas have been applied to classify active matter \cite{dulaney2021machine,xue2022machine}. In particular, to achieve this, ref. \cite{dulaney2021machine} has applied machine learning techniques to determine in which phase an active test particle is based on the particles in the vicinity of that particle. This has been discussed in the context of motility-induced phase separation of active particles. 
A different application of machine learning techniques to analyze phase diagrams has been reported in ref. \cite{jeckel2019learning} where learning techniques have been used to 
identify different sources for bacterial swarming based on experimental and simulation data \cite{jeckel2019learning}. For this complex system, the full space-time phase diagram of bacterial swarm expansion was analyzed and classified by machine learning techniques.
\\Another popular and highly relevant application of machine learning methods to active matter concerns the learning of governing equations describing active matter ~\cite{Supekar_2023, maddu2022learning}. Such methods use data (e.g. 
particle trajectories) from experiments or stochastic particle based simulations to predict 
governing equations describing the system at coarse grained scales. 
The enormous potential of such data-driven approaches has already been demonstrated in recent works that exploit techniques like symbolic regression, autoencoder networks, and sparse regression \cite{brunton2016discovering,rudy2017data,Champion_2019, Supekar_2023}.
Exemplaric applications to active matter have recently been proposed in 
refs.~\cite{maddu2022learning} and \cite{Supekar_2023}. Such methods are 
highly useful to find new phenomenological hydrodynamic descriptions of fluctuating active matter and beyond. See Fig.~\ref{fig:supekarDunkelPnas2023} for a very recent example. Here not the strategies of the many-body system are learned but the underlying governing equations.

\section{Conclusions and outlook}
In conclusion, micron-sized active particles can be equipped with different levels of ``intelligence''. In particular it is now possible to dynamically control them with external feedback control systems which can be combined with suitable machine learning approaches. This results in ``smart'' active particles that can sense and react to their environment, communicate with each other and learn from past experience (although the actual learning process takes place in a computer). 
However, as these particles depend on their external control system and on the external learning platform they do not yet reach the autonomy of biological active particles like bacteria or other microorganisms. 
In fact, 
equipping synthetic agents with sensors, actuators and processors to make them smart and autonomous remains as a persistent challenge at the micro-scale. (At the macroscale such an approach leads to electronic robots, which are many orders of magnitude larger than active colloidal particles.) 
\\Today we are much closer to the vision of emergent intelligence formulated by Hillis more than 30 years ago \cite{hillis1988intelligence}. Overall a flourishing future for applications of artificial intelligence to active particles is lying ahead. While most of the current work is theoretical or based on simulations, the experimental realization of autonomous and intelligent micron-sized particles remains highly challenging. 

\section*{Acknowledgments}

The work of HL was supported by the German Research Foundation (DFG) via project LO 418/29-1 (project number 522595197).


\begin{thebibliography}{10}
\providecommand{\url}[1]{{#1}}
\providecommand{\urlprefix}{URL }
\expandafter\ifx\csname urlstyle\endcsname\relax
  \providecommand{\doi}[1]{DOI \discretionary{}{}{}#1}\else
  \providecommand{\doi}{DOI \discretionary{}{}{}\begingroup
  \urlstyle{rm}\Url}\fi

\bibitem{mestre2022colloidal}
F.~Sagués~Mestre, \emph{Colloidal Active Matter: Concepts, Experimental
  Realizations, and Models} (CRC Press, 2022)

\bibitem{cichos2020machine}
F.~Cichos, K.~Gustavsson, B.~Mehlig, G.~Volpe, Nat. Mach. Intell. \textbf{2},
  94 (2020)

\bibitem{bechinger2016active}
C.~Bechinger, R.~Di~Leonardo, H.~L{\"o}wen, C.~Reichhardt, G.~Volpe, G.~Volpe,
  Rev. Mod. Phys. \textbf{88}, 045006 (2016)

\bibitem{gompper20202020}
G.~Gompper, R.G. Winkler, T.~Speck, A.~Solon, C.~Nardini, F.~Peruani,
  H.~L{\"o}wen, R.~Golestanian, U.B. Kaupp, L.~Alvarez, J. Phys. Condens.
  Matter \textbf{32}, 193001 (2020)

\bibitem{paxton2004catalytic}
W.F. Paxton, K.C. Kistler, C.C. Olmeda, A.~Sen, S.K. St.~Angelo, Y.~Cao, T.E.
  Mallouk, P.E. Lammert, V.H. Crespi, J. Am. Chem. Soc. \textbf{126}, 13424
  (2004)

\bibitem{nosenko2020active}
V.~Nosenko, F.~Luoni, A.~Kaouk, M.~Rubin-Zuzic, H.~Thomas, Phys. Rev. Res.
  \textbf{2}, 033226 (2020)

\bibitem{nasiri2023optimal}
M.~Nasiri, H.~L{\"o}wen, B.~Liebchen, EPL \textbf{142}, 17001 (2023)

\bibitem{cichos2023artificial}
F.~Cichos, S.M. Landin, R.~Pradip, in \emph{Intelligent Nanotechnology}
  (Elsevier, 2023), p. 113

\bibitem{yethiraj2007tunable}
A.~Yethiraj, Soft Matter \textbf{3}, 1099 (2007)

\bibitem{lowen2013introduction}
H.~L{\"o}wen, The European Physical Journal Special Topics \textbf{222}, 2727
  (2013)

\bibitem{liebchen2018synthetic}
B.~Liebchen, H.~Löwen, Acc. Chem. Res. \textbf{51}, 2982 (2018)

\bibitem{lavergne2019group}
F.A. Lavergne, H.~Wendehenne, T.~B{\"a}uerle, C.~Bechinger, Science
  \textbf{364}, 70 (2019)

\bibitem{khadka2018active}
U.~Khadka, V.~Holubec, H.~Yang, F.~Cichos, Nat. Commun. \textbf{9}, 3864 (2018)

\bibitem{bauerle2018quorum}
T.~B{\"a}uerle, A.~Fischer, T.~Speck, C.~Bechinger, Nat. Commun. \textbf{9},
  3232 (2018)

\bibitem{sprenger2020active}
A.R. Sprenger, M.A. Fernandez-Rodriguez, L.~Alvarez, L.~Isa, R.~Wittkowski,
  H.~L{\"o}wen, Langmuir \textbf{36}, 7066 (2020)

\bibitem{mijalkov2016engineering}
M.~Mijalkov, A.~McDaniel, J.~Wehr, G.~Volpe, Phys. Rev. X \textbf{6}, 011008
  (2016)

\bibitem{patra2013intelligent}
D.~Patra, S.~Sengupta, W.~Duan, H.~Zhang, R.~Pavlick, A.~Sen, Nanoscale
  \textbf{5}(4), 1273 (2013)

\bibitem{alapan2018soft}
Y.~Alapan, O.~Yasa, O.~Schauer, J.~Giltinan, A.F. Tabak, V.~Sourjik, M.~Sitti,
  Sci. Robot. \textbf{3}, eaar4423 (2018)

\bibitem{gu2022artificial}
H.~Gu, E.~Hanedan, Q.~Boehler, T.Y. Huang, A.J. Mathijssen, B.J. Nelson, Nat.
  Mach. Intell. \textbf{4}, 678 (2022)

\bibitem{spehr2003identification}
M.~Spehr, G.~Gisselmann, A.~Poplawski, J.A. Riffell, C.H. Wetzel, R.K. Zimmer,
  H.~Hatt, Science \textbf{299}, 2054 (2003)

\bibitem{eisenbach2006sperm}
M.~Eisenbach, L.C. Giojalas, Nature reviews Molecular cell biology
  \textbf{7}(4), 276 (2006)

\bibitem{tsang2020roads}
A.C. Tsang, E.~Demir, Y.~Ding, O.S. Pak, Adv. Intell. Syst. \textbf{2}, 1900137
  (2020)

\bibitem{najafi2004simple}
A.~Najafi, R.~Golestanian, Phys. Rev. E \textbf{69}, 062901 (2004)

\bibitem{avron2005pushmepullyou}
J.E. Avron, O.~Kenneth, D.H. Oaknin, New J. Phys. \textbf{7}, 234 (2005)

\bibitem{ledesma2012circle}
R.~Ledesma-Aguilar, H.~L{\"o}wen, J.M. Yeomans, Eur. Phys. J. E \textbf{35},
  9746 (2012)

\bibitem{daddi2018swimming}
A.~Daddi-Moussa-Ider, M.~Lisicki, C.~Hoell, H.~L{\"o}wen, J. Chem. Phys.
  \textbf{148}, 134904 (2018)

\bibitem{tsang2020self}
A.C.H. Tsang, P.W. Tong, S.~Nallan, O.S. Pak, Phys. Rev. Fluids \textbf{5},
  074101 (2020)

\bibitem{zou2022gait}
Z.~Zou, Y.~Liu, Y.N. Young, O.S. Pak, A.C. Tsang, Commun. Phys. \textbf{5}, 158
  (2022)

\bibitem{abdi2023self}
H.~Abdi, H.N. Pishkenari, Eng. Appl. Artif. Intell. \textbf{123}, 106188 (2023)

\bibitem{qin2023reinforcement}
K.~Qin, Z.~Zou, L.~Zhu, O.S. Pak, Phys. Fluids \textbf{35}, 032003 (2023)

\bibitem{liu2021mechanical}
Y.~Liu, Z.~Zou, A.C.H. Tsang, O.S. Pak, Y.N. Young, Phys. Fluids \textbf{33},
  062007 (2021)

\bibitem{kaur2023adaptive}
H.~Kaur, T.~Franosch, M.~Caraglio, Mach. Learn.: Sci. Technol. \textbf{4},
  035008 (2023)

\bibitem{goh2022noisy}
S.~Goh, R.G. Winkler, G.~Gompper, New J. Phys. \textbf{24}, 093039 (2022)

\bibitem{hartl2021microswimmers}
B.~Hartl, M.~H{\"u}bl, G.~Kahl, A.~Z{\"o}ttl, Proc. Natl. Acad. Sci.
  \textbf{118}, e2019683118 (2021)

\bibitem{ramakrishnan2023learning}
R.O. Ramakrishnan, B.M. Friedrich, EPL \textbf{142}, 47001 (2023)

\bibitem{liebchen2019optimal}
B.~Liebchen, H.~L\"owen, EPL \textbf{127}, 34003 (2019)

\bibitem{daddi2021hydrodynamics}
A.~Daddi-Moussa-Ider, H.~L{\"o}wen, B.~Liebchen, Commun. Phys. \textbf{4}, 15
  (2021)

\bibitem{sutton2018reinforcement}
R.S. Sutton, A.G. Barto, \emph{Reinforcement learning: An introduction} (MIT
  press, 2018)

\bibitem{Yang2018}
Y.~Yang, M.A. Bevan, ACS Nano \textbf{12}, 10712 (2018)

\bibitem{Yang2019}
Y.~Yang, M.A. Bevan, B.~Li, Adv. Intell. Syst. \textbf{2}, 1900106 (2020)

\bibitem{biferale2019}
L.~Biferale, F.~Bonaccorso, M.~Buzzicotti, P.~Clark Di~Leoni, K.~Gustavsson,
  Chaos \textbf{29}, 103138 (2019)

\bibitem{yang2020micro}
Y.~Yang, M.A. Bevan, B.~Li, Adv. Theory Simul. \textbf{3}, 2000034 (2020)

\bibitem{xu2021brownian}
K.~Xu, Y.~Yang, B.~Li, Adv. Intell. Syst. \textbf{3}, 2100115 (2021)

\bibitem{alageshan2020machine}
J.K. Alageshan, A.K. Verma, J.~Bec, R.~Pandit, Physical Review E
  \textbf{101}(4), 043110 (2020)

\bibitem{zhu2022numerical}
Y.~Zhu, J.H. Pang, Proc. Inst. Mech. Eng. C J. Mech. Eng. Sci. \textbf{237},
  2450 (2023)

\bibitem{zhu2022point}
Y.~Zhu, J.H. Pang, F.B. Tian, Front. Phys. \textbf{10}, 237 (2022)

\bibitem{nasiri2022reinforcement}
M.~Nasiri, B.~Liebchen, New J. Phys. \textbf{24}, 073042 (2022)

\bibitem{muinos2021reinforcement}
S.~Mui{\~n}os-Landin, A.~Fischer, V.~Holubec, F.~Cichos, Science Robotics
  \textbf{6}(52), eabd9285 (2021)

\bibitem{schneider2019}
E.~Schneider, H.~Stark, EPL \textbf{127}, 64003 (2019)

\bibitem{gunnarson2021learning}
P.~Gunnarson, I.~Mandralis, G.~Novati, P.~Koumoutsakos, J.O. Dabiri, Nat.
  Commun. \textbf{12}, 7143 (2021)

\bibitem{lozano2016phototaxis}
C.~Lozano, B.~Ten~Hagen, H.~L{\"o}wen, C.~Bechinger, Nat. Commun.
  \textbf{7}(1), 12828 (2016)

\bibitem{monderkamp2022active}
P.A. Monderkamp, F.J. Schwarzendahl, M.A. Klatt, H.~L{\"o}wen, Mach. Learn.:
  Sci. Technol. \textbf{3}, 045024 (2022)

\bibitem{dijkstra1959note}
E.W. Dijkstra, Numerische Mathematik \textbf{1}, 269 (1959)

\bibitem{meredith2020predator}
C.H. Meredith, P.G. Moerman, J.~Groenewold, Y.J. Chiu, W.K. Kegel, A.~van
  Blaaderen, L.D. Zarzar, Nat. Chem. \textbf{12}, 1136 (2020)

\bibitem{sengupta2011chemotactic}
A.~Sengupta, T.~Kruppa, H.~L{\"o}wen, Phys. Rev. E \textbf{83}, 031914 (2011)

\bibitem{schwarzendahl2021barrier}
F.J. Schwarzendahl, H.~L{\"o}wen, EPL \textbf{134}, 48005 (2021)

\bibitem{liebchen2020modeling}
B.~Liebchen, H.~L{\"o}wen, in \emph{Chemical Kinetics: Beyond the Textbook}
  (World Scientific, 2020), p. 493

\bibitem{wang2020reinforcement}
X.~Wang, J.~Cheng, L.~Wang, Ecol. Complex. \textbf{42}, 100815 (2020)

\bibitem{gerhard2021hunting}
M.~Gerhard, A.~Jayaram, A.~Fischer, T.~Speck, Phys. Rev. E \textbf{104}, 054614
  (2021)

\bibitem{janosov2017group}
M.~Janosov, C.~Vir{\'a}gh, G.~V{\'a}s{\'a}rhelyi, T.~Vicsek, New J. Phys.
  \textbf{19}, 053003 (2017)

\bibitem{ranganathan2022optimal}
A.~Ranganathan, A.~Heyde, A.~Gupta, L.~Mahadevan, arXiv preprint
  arXiv:2211.04352  (2022)

\bibitem{zampetaki2021collective}
A.V. Zampetaki, B.~Liebchen, A.V. Ivlev, H.~L{\"o}wen, Proc. Natl. Acad. Sci.
  \textbf{118}, e2111142118 (2021)

\bibitem{ziepke2022multi}
A.~Ziepke, I.~Maryshev, I.S. Aranson, E.~Frey, Nat. Commun. \textbf{13}, 6727
  (2022)

\bibitem{panait2005cooperative}
L.~Panait, S.~Luke, Auton. Agents Multi-Agent Syst. \textbf{11}, 387 (2005)

\bibitem{durve2020learning}
M.~Durve, F.~Peruani, A.~Celani, Phys. Rev. E \textbf{102}, 012601 (2020)

\bibitem{Speck2014}
T.~Speck, J.~Bialk{\'e}, A.M. Menzel, H.~L{\"o}wen, Phys. Rev. Lett.
  \textbf{112}, 218304 (2014)

\bibitem{tovey2023environmental}
S.~Tovey, D.~Zimmer, C.~Lohrmann, T.~Merkt, S.~Koppenhoefer, V.L. Heuthe,
  C.~Bechinger, C.~Holm, arXiv preprint arXiv:2307.00994  (2023)

\bibitem{verma2018efficient}
S.~Verma, G.~Novati, P.~Koumoutsakos, Proc. Natl. Acad. Sci. \textbf{115}, 5849
  (2018)

\bibitem{GrauerPREPoptimizing}
J.~Grauer, H.~Löwen, F.~Schwarzendahl, B.~Liebchen, Preprint, in preparation
  (2023)

\bibitem{Supekar_2023}
R.~Supekar, B.~Song, A.~Hastewell, G.P. Choi, A.~Mietke, J.~Dunkel, Proc. Natl.
  Acad. Sci. \textbf{120}, e2206994120 (2023)

\bibitem{carrasquilla2017machine}
J.~Carrasquilla, R.G. Melko, Nat. Phys. \textbf{13}, 431 (2017)

\bibitem{rodriguez2019identifying}
J.F. Rodriguez-Nieva, M.S. Scheurer, Nat. Phys. \textbf{15}, 790 (2019)

\bibitem{wetzel2017unsupervised}
S.J. Wetzel, Phys. Rev. E \textbf{96}, 022140 (2017)

\bibitem{van2017learning}
E.P. Van~Nieuwenburg, Y.H. Liu, S.D. Huber, Nat. Phys. \textbf{13}, 435 (2017)

\bibitem{dulaney2021machine}
A.R. Dulaney, J.F. Brady, Soft Matter \textbf{17}, 6808 (2021)

\bibitem{xue2022machine}
T.~Xue, X.~Li, X.~Chen, L.~Chen, Z.~Han, arXiv preprint arXiv:2210.00161
  (2022)

\bibitem{jeckel2019learning}
H.~Jeckel, E.~Jelli, R.~Hartmann, P.K. Singh, R.~Mok, J.F. Totz, L.~Vidakovic,
  B.~Eckhardt, J.~Dunkel, K.~Drescher, Proc. Natl. Acad. Sci. \textbf{116},
  1489 (2019)

\bibitem{maddu2022learning}
S.~Maddu, Q.~Vagne, I.F. Sbalzarini, arXiv preprint arXiv:2201.08623  (2022)

\bibitem{brunton2016discovering}
S.L. Brunton, J.L. Proctor, J.N. Kutz, Proc. Natl. Acad. Sci. \textbf{113},
  3932 (2016)

\bibitem{rudy2017data}
S.H. Rudy, S.L. Brunton, J.L. Proctor, J.N. Kutz, Science Advances \textbf{3},
  e1602614 (2017)

\bibitem{Champion_2019}
K.~Champion, B.~Lusch, J.N. Kutz, S.L. Brunton, Proc. Natl. Acad. Sci.
  \textbf{116}, 22445 (2019)

\bibitem{hillis1988intelligence}
W.D. Hillis, Daedalus \textbf{117}, 175 (1988)

\end{thebibliography}

\end{document}